\documentclass{webofc}
\usepackage[varg]{txfonts}   
\usepackage{hyperref}
\usepackage{comment}
\begin{document}
\title{ Anomalies in the heavy sector}
%
%

\author{\firstname{Giulia} \lastname{Ricciardi}\inst{1,2}\fnsep\thanks{\email{giulia.ricciardi2@unina.it}} }

\institute{Dipartimento di Fisica E. Pancini, Universit\`a  di Napoli Federico II \\
  Complesso Universitario di Monte Sant'Angelo, Via Cintia,
I-80126 Napoli, Italy
\and
          INFN, Sezione di Napoli,
  Complesso Universitario di Monte Sant'Angelo, Via Cintia,
I-80126 Napoli, Italy
}

\abstract{%
Heavy flavour physics is well  described  by the Standard Model with the exception of some measurements which could be pointing to new physics. We briefly review status and prospects of  $R(D^{(\ast)})$ and $R(K^{(\ast)})$   anomalies and
 also address  the long standing tension in the $|V_{cb}|$ and $|V_{ub}|$ exclusive/inclusive determinations.}

\maketitle
\section{Introduction}
\label{intro}
 In a general scenario of optimal agreement within the Standard Model (SM),   the heavy flavour physics sector exhibits some measurements  which present non-significant but intriguing  tensions with SM predictions. Here we briefly address  the discrepancy between  measured ratios in  semileptonic $B$ decays and the corresponding SM predictions, which
 may hint at lepton flavour non-universality  or lepton flavour violation.

Semileptonic $B$ decays are also the processes of election when it comes to a precise determination of the parameters $|V_{xb}|$  of the Cabibbo-Kobayashi-Maskawa (CKM) matrix. 
Their precise measurement allows a powerful check of the unitarity of the CKM matrix. However, there is  long-standing tension among the $|V_{xb}|$ values, depending on whether they  are  extracted  using exclusive or inclusive semi-leptonic $B$ decays. We briefly review  recent progress on  $|V_{xb}|$ determinations, prompted by  the larger and larger  samples of  $B$ mesons collected at the $B$-factories and at LHCb, and by  concomitant progress in theoretical computations~\footnote{For brief overviews see for example \cite{Ricciardi:2014aya, Ricciardi:2016pmh, Ricciardi:2017lne,Koppenburg:2017mad, Ricciardi:2019zph} and references therein.}.

\section{ $R(D)$ and $R(D^{\ast})$}

 Exclusive semi-tauonic $B$ decays were
first observed by the Belle Collaboration in 2007 \cite{Matyja:2007kt}.
Subsequent
analyses by Babar and Belle \cite{Aubert:2007dsa, Bozek:2010xy,Huschle:2015rga} measured
 branching fractions above, although consistent with, the SM predictions.
The ratio of branching fractions (the denominator is the average for $l \in \{e, \mu\}$)
\begin{equation}
R(D^{(\ast)}) \equiv  \frac{{\cal{B}}( B \to D^{(\ast)} \tau \nu_\tau)}{{\cal{B}}( B \to D^{(\ast)} l  \nu_l)}
\label{ratiotau0}
\end{equation}
 is typically used instead of the absolute branching fraction
of $ B \to D^{(\ast)} \tau  \nu_\tau$ decays to cancel  uncertainties common to the numerator and the denominator.
These include the CKM matrix element $|V_{cb}|$ and several theoretical uncertainties on hadronic form factors and experimental reconstruction effects. 
Since  the strong force does not couple directly to leptons, its
 effect on the semileptonic decays in different lepton families is identical, aside from the difference in the phase space. The
ratio between the branching fractions of these decays is therefore predicted with O(1\%)
precision.
 In 2012-2013
Babar
 has measured
$R(D^{(\ast)}) $ by using  its full data sample \cite{Lees:2012xj, Lees:2013uzd},
and reported a significant excess over the SM expectation, confirmed  in 2015
by LHCb \cite{Aaij:2015yra}.
In 2016 such excess has  been confirmed also by the Belle collaboration, which has performed the first measurement of $R(D^{\ast})$
using the semileptonic tagging method, giving \cite{Belle:2016kgw}
\begin{equation}
R(D^{\ast})  = 0.302\pm 0.030\pm 0.011
\label{ratiotauBlle2016}
\end{equation}
where the first error is statistic and the second one is systematic.
%
By selecting the previous measurements in the time-span  2012-2016 \cite{Lees:2012xj, Lees:2013uzd,Aaij:2015yra, Huschle:2015rga, Belle:2016kgw}, in 2016 the HFLAV Collaboration found the average \cite{HFAG2016}
\begin{equation}
R(D)  = 0.397 \pm 0.040 \pm 0.028  \qquad \qquad
R(D^{\ast})= 0.316 \pm 0.016 \pm 0.010
\label{ratiotau}
\end{equation}
where the first uncertainty is statistical and the second is
systematic. $R(D) $ and $R(D^{\ast})$  exceed the SM
 value $R(D) ^{SM} = 0.300\pm 0.008 $, given by the HPQCD collaboration \cite{Na:2015kha}, and the SM value $R(D^{\ast})^{SM}= 0.252\pm 0.003$ \cite{Fajfer:2012vx}
  by  1.9$\sigma$ and 3.3$\sigma$, respectively.
The combined analysis of  $R(D) $ and  $R(D^{\ast})$, taking
into account measurement correlations, finds that the deviation
is 4$\sigma$ from the SM prediction.

Fast forward to 2022, the HFLAV analysis  has added to the previous results from Babar, Belle and LHCb \cite{Lees:2012xj, Lees:2013uzd,Huschle:2015rga,Aaij:2015yra} the most recent results from Belle  \cite{Belle:2016dyj, Belle:2017ilt,Belle:2019rba} and LHCb \cite{LHCb:2017smo,LHCb:2017rln, LHCbpreli22}. The HFLAV determination finds lower average values and smaller errors, namely \cite{Amhis:2022mac}  
\begin{equation}
R(D)   = 0.358 \pm 0.025 \pm 0.012  \qquad \qquad
R(D^{\ast})= 0.285 \pm 0.010 \pm 0.008
\label{ratiotau25}
\end{equation}
New SM calculations have become available since the 2016. The results are compatible with the previous ones, although generally more accurate or robust. The HFLAV has performed their arithmetic average resulting in \cite{Amhis:2022mac}  $R(D) ^{SM} = 0.298\pm 0.004 $ and $R(D^{\ast})^{SM}= 0.254\pm 0.005$.
The experimental averages \eqref{ratiotau25}
  exceed these SM
 values   by  2.16$\sigma$ and 2.26$\sigma$, respectively.
  The combined analysis of  $R(D) $ and  $R(D^{\ast})$ finds that the deviation
is 3.2$\sigma$ from the SM prediction.


The actual recorded luminosity at Belle II is 428 fb$^{-1}$, to be compared with  Belle (988 fb$^{-1}$) and  BaBar (513 fb$^{-1}$);
the ultimate goal is to reach 50 ab$^{-1}$.  Investigating the anomaly through precision measurements of
$R(D^{(\ast)})$ is a chief goal of Belle II.  
%
At Belle II a better understanding of
backgrounds tails under the signal  and a reduction of the uncertainty  to 3\% for $R(D^{\ast})$  and 5\% for   $R(D)$ is expected at 5 ab$^{-1}$\cite{Belle-II:2022cgf}.

Belle II has the unique potential to  measure the ratio $R(X)$ of the inclusive
rate $B \to  X \tau  \nu$ to the lower-mass lepton counterparts. This ratio would probe both electron
and muon modes with a precise consistency check whose phenomenological interpretation
is independent from LQCD uncertainties that affect the other observables. However,
this is a challenging measurement attempted and never completed at previous B-factory
experiments. The main experimental challenge is to control the significant systematic
uncertainties associated with background composition \cite{Belle-II:2022cgf}. 
Belle II is preparing and this year it has  measured the ratio $R(X_{e/\mu})$  of the inclusive
rate $B \to  X_e  \nu$ to $B \to  X_\mu  \nu$, which is in good agreement with the SM \cite{Junkerkalefeld:3125}.

The $B$-factories have performed the most
precise measurements of $R(D)$ and $R(D^\ast)$ to date thanks to their ability to significantly
constrain the kinematics of these neutrinos by leveraging their knowledge of the $e^+ e^-$
collision energy. LHCb, however, is expected to surpass the B-factories precision and
reach uncertainties down to the percent level from the analysis of the enormous
data samples expected from the operation of the Upgrades I and II detectors, provided
that the relevant systematic uncertainties can be properly controlled \cite{LHCb:2022ine}.

Other ratios are currently under investigation.
 LHCb has found the value \cite{LHCb:2017vlu}  $R(J/\Psi)= 0.71 \pm 0.17 \pm 0.18$, which is 
 2.1$\sigma$ from the LQCD value \cite{Harrison:2020nrv}
 $R(J/\Psi)^{\mathrm{HPQCD}}=0.2582\pm 0.0038$. 
 Another LHCb measurement gives \cite{LHCb:2022piu}
 \begin{equation}
 R(\Lambda_c) \equiv  \frac{{\cal{B}}( \Lambda_b \to \Lambda_c \tau \nu_\tau)}{{\cal{B}}( \Lambda_c \to \Lambda_c \mu  \nu_\mu)}=
 0.242 \pm 0.026 \pm 0.040 \pm 0.059
\label{ratiotau06}
  \end{equation}
where the external branching fraction uncertainty
from the channel $\Lambda^0_b \to \Lambda^+_c \mu^-  \bar \nu_\mu$
contributes to the last term. The last ratio seems can  be related to the values of  $R(D)$ and $R(D^*)$ \cite{Blanke:2018yud}.

\section{$R(K)$ and $R(K^*)$}

While $R(B)$ is defined as the ratio of branching fractions of decays that occur  at tree level in the SM at the lowest perturbative order, the observable  $R(K)$ and $R(K^*)$ are defined as the ratios of branching fractions of rare decays. 
In general, for semileptonic decays of $B$ hadrons,  where $B = B^+, B^0, B^0_s$ or $\Lambda^0_b$, the ratio $R(H)$     is defined in the dilepton mass-squared range $q^2_{\mathrm{min}} < q^2 < q^2_{\mathrm{max}}$ as
\begin{equation}
 R(H)=\frac{\int_{q^2_{\mathrm{min}}}^{ q^2_{\mathrm{max}}} \frac {d {\cal{B}}( B \to H\mu^+ \mu^-)}{dq^2}{ \tiny dq^2}}{\int_{q^2_{\mathrm{min}}}^{ q^2_{\mathrm{max}}} \frac {d {\cal{B}}( B \to H e^+ e^-)}{dq^2} dq^2}
\end{equation}
where $H$ can be an hadron as  $K$ or a
combination of particles such as a proton and charged kaon, $pK^-$.

LHCb has provided the most precise  measurements of the branching fractions of the $ B^+ \to K^{(\ast)+}   \mu^+ \mu^-$ and  $ B^0 \to K^{(\ast)0}   \mu^+ \mu^-$   decays \cite{LHCb:2014cxe, LHCb:2012juf, LHCb:2013zuf}.
All these measurements are below the
SM prediction.
At LHCb,  $R(K)$ has been measured to be \cite{Aaij:2014ora}
\begin{equation}
R(K^+) =0.745^{+0.090}_{-0.074}\pm 0.036
\label{ratiotau0}
\end{equation}
where the first error is statistical and the second one is systematic. The measurement was performed
across the dilepton
invariant-mass-squared range [1,6] GeV$^2$.
This result is 2.6$\sigma$ deviations away from the SM prediction $R(K)^{SM}=1.0003 \pm 0.0001$
\cite{Bobeth:2007dw}. The impact of radiative corrections  has been estimated not to exceed a few \%
 \cite{Bordone:2016gaq}.
The  result in Eq. \eqref{ratiotau0} has been superseded in 2021 by an LHCB analysis which uses  essentially identical techniques but an additional 4 fb$^{-1}$
of data collected in 2017 and 2018 \cite{LHCb:2021trn}. Hence, the entire amount of data was recorded during the years 2011, 2012 and 2015–2018, in which
the centre-of-mass energy of the collisions was 7, 8 and 13 TeV, and correspond to an
integrated luminosity of 9 fb$^{-1}$. The analysis \cite{LHCb:2021trn} gives, in the dilepton
invariant-mass-squared range [1,6] GeV$^2$, the value 
\begin{equation}
R(K^+)  =0.846^{+0.042\, +0.013}_{-0.039\, -0.012} 
\label{eq:ratiotau1}
\end{equation}
where the first uncertainty is statistical and the second systematic. Combining the
uncertainties they obtain $ R(K^+)  =0.846^{+0.044}_{-0.041}$. 
There are previous measurements by Babar \cite{BaBar:2012mrf} and Belle \cite{BELLE:2019xld}, but the LHCb measurement
is the most precise  to date and is
consistent with the SM expectation at the level of 3.1 standard
deviations.

\begin{figure}[h]
\centering
\includegraphics[width=9cm,clip]{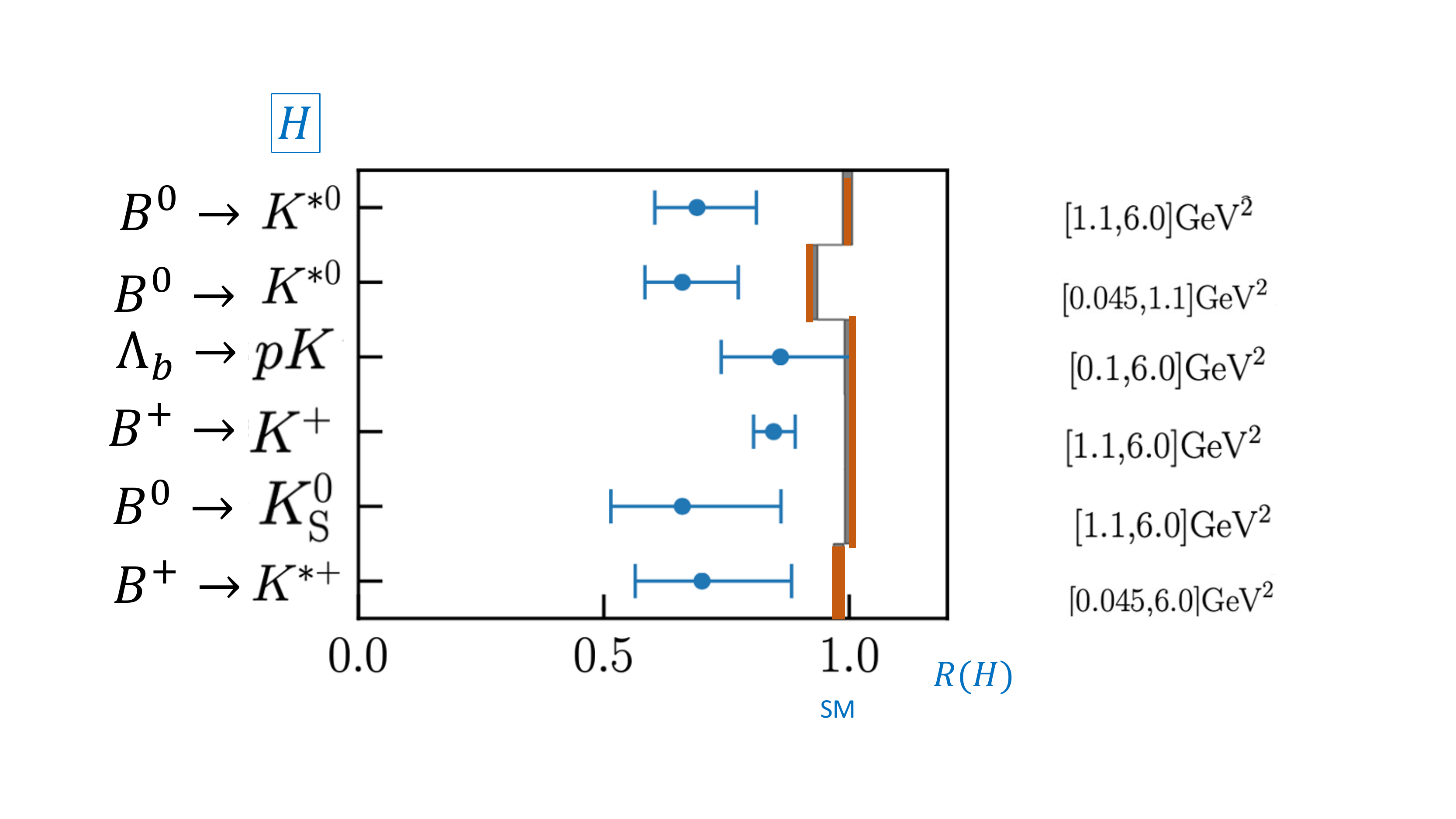}
\caption{Current LHCb values for semileptonic ratios $R(H)$, where the decaying particle is also indicated. The first two results are taken from  \cite{LHCb:2017avl}, the baryon decay has been analyzed in \cite{ LHCb:2019efc}, $R(K^+)$ in     \cite{LHCb:2021trn}, and the last two measurements in \cite{LHCb:2021lvy}. }
\label{fig-1}       
\end{figure}

The other ratio measurements from LHCb include $R(K^{\ast 0}) $ with the
Run 1 data \cite{LHCb:2017avl}  
in two regions of the dilepton invariant mass squared; the value found for the interval 
[0.045,1.1] GeV$^2$ is
\begin{equation}
R(K^{\ast 0})  =0.66^{+0.11}_{-0.07} \pm 0.03
\label{eq:ratiotau12}
\end{equation}
and the value found  for the interval 
[1.1,6.0] GeV$^2$ is
\begin{equation}
R(K^{\ast 0})  =0.69^{+0.11}_{-0.07} \pm 0.05
\label{eq:ratiotau12}
\end{equation}
These results, which represent the most precise measurements of $R(K^{\ast 0}) $ to date, are compatible with
the Standard Model expectations at the level of 2.1-2.3 and 2.4-2.5 standard deviations (depending on the theoretical predictions) in
the two $q^2$ regions. Previous results from Babar \cite{BaBar:2012mrf} and Belle \cite{Belle:2009zue} are also available.
LHCb current results are reported in figure~\ref{fig-1}, where also the relevant dilepton invariant mass squared interval is indicated.

%

\section{Data interpretation}

The alleged   breaking of lepton-flavour universality suggested by most of the data is quite large,  and several theoretical models have been tested against the experimental results.
A model independent approach, where the SM is considered as an effective low-energy theory,  is mostly used. The SM Effective Theory (SMEFT) is the effective theory above the electroweak (EW) scale, incorporating
the full SM content, with non renormalizable operators which are gauge-invariant under the SM gauge group. 
The main assumption inherent to the SMEFT framework is that new physics interactions have
been integrated out at some high scale. The SMEFT aims at characterizing experimental deviations from the predictions of the Standard Model and pointing towards the structures of its possible extensions
 above the EW scale.
The Weak Effective Theory (WET) is the effective theory valid below the EW scale which includes all SM particles
except the the gauge bosons, the Higgs and the $t$ quark. In contrast to the SMEFT case, the dimension-six operators in WET are not invariant under the full SM gauge group, but only under $SU(3)_c\times U(1)_{em}$, as this EFT is defined below the EW scale where 
$SU(2)_L\times U(1)_{Y}$ is already broken. For a given set of SMEFT dimension-six operators with the corresponding Wilson coefficients specified at the higher scale, one has to perform the renormalization group evolution  of the SMEFT Wilson coefficients  down to the EW scale,  match the given set of SMEFT operators onto the WET ones at the EW scale and finally perform the RGE of the WET Wilson coefficients from the EW down to the dynamical scale of the $B$ meson. This way, one can then bridge the gap between the SMEFT Lagrangian and the low-energy measurements in $B$ physics. 

New physics effects in semiletonic decays into light quarks  are stringently constrained. For instance the ratio \cite{Belle:2018ezy}
\begin{equation}
 \frac{{\cal{B}}( B^0 \to D^{\ast -} e^+  \nu_e )}{{\cal{B}}(  B^0 \to D^{\ast -} \mu^+ \nu_\mu )}=1.01 \pm 0.01\pm 0.03
 \end{equation}
is in agreement with the SM. Hence new physics effects
are generally assumed in semitauonic $B$ decays.
The relevant terms of the WET
Lagrangian for the underlying $b\to c\tau\nu$ transition are~\cite{ Aebischer:2015fzz, Aebischer:2017gaw,Tanaka:2012nw} 
\begin{equation}
\mathcal{L}_{\text{eff}} = -\frac{4 G_F}{\sqrt{2}} V^{}_{cb} \Big[(1+C_{V}^{L}) O_{V}^L +   C_{S}^{R} O_{S}^{R} 
   +C_{S}^{L} O_{S}^L+   C_{T} O_{T}\Big] +\text{h.c.}
\end{equation}
where $G_F$ is the Fermi constant, 
$ V^{}_{cb}$ is an element of the CKM matrix
and the dimension-six four-fermion operators are
\begin{align*}
  & O_{V}^L  = \left(\bar c\gamma ^{\mu } P_L b\right)  \left(\bar\tau \gamma_{\mu } P_L \nu_{\tau}\right)\,, &&    O_{S}^R  = \left( \bar c P_R b \right) \left( \bar\tau P_L \nu_{\tau}\right)\,,\\
  & O_{T}  = \left( \bar c \sigma^{\mu\nu}P_L  b \right) \left( \bar\tau \sigma_{\mu\nu} P_L \nu_{\tau}\right)\,, &&
   O_{S}^L  = \left( \bar c P_L b \right) \left( \bar\tau P_L \nu_{\tau}\right)  \,.
\end{align*}

Non-zero Wilson coefficients relate to the possible tree-level NP mediators:
(i)
 exchange of an heavy $W^\prime$ boson with left-handed couplings. Simplified models (spin-1 colorless weak triplet, two Higgs doublet models, a spin-0 or spin-1 leptoquark) are in tension with existing $\tau^+\tau^-$ LHC results \cite{Faroughy:2016osc}.
These constraints do no apply to the low mass region in the $W^\prime$ and vector leptoquark models; (ii)
 exchange of a charged Higgs boson. This scenario is in tension with the LHC mono-$\tau$ data and it can be related to (large) branching ratios for $B_c \to \tau \nu$ decays. (iii)
  leptoquark scenarios. 
They generally evade the LHC mono-$\tau$ tests, but  they are subject to LHC constraints from their pair-production and t-channel mediated di-lepton processes.
Other constraints may come from characteristics of the models, as for instance induced CP violations or the presence of colour-octet resonances, often introduced together in UV-complete models.
In general, and in all the above models, more severe constraints can appear in concrete UV completions.

The $b\to s\ell^+\ell^-$ transitions are described by WET Lagrangian
\begin{equation}
\mathcal{L}_{\text{eff}}=\frac{4 G_F}{\sqrt{2}} V_{tb}^* V_{ts} \frac{e^2}{16\pi^2}\sum_i(C_i  O_i +C'_i {O}'_i)+\text{h.c.}\,,
\end{equation}
where the operators most sensitive to new physics are:
\begin{equation}
O_{7}^{(\prime)}=\frac{m_b}{e}(\bar s\sigma_{\mu\nu} P_{R(L)} b)F^{\mu\nu} \quad O_{9,10}^{(\prime)} = (\bar sb)_{V\mp A} (\bar\ell\ell)_{V,A}
\end{equation}
and their corresponding Wilson coefficients 
have contributions from the SM processes as well as from  new physics ones.
Several global fits have been performed in the
literature (see for instance ~\cite{Capdevila:2017bsm,Celis:2017doq,Alok:2017sui,Camargo-Molina:2018cwu,Datta:2019zca,Aebischer:2019mlg,Aoude:2020dwv}). 
They can be fully data driven, which means that no
assumption is done about charming
penguins, or partly or fully model
dependent, assuming assume LCSR
results for charming
penguins.
There is a general agreement among the fits, which 
show that even the
simple one-dimensional new physics scenarios, such as $ C_9^{\mathrm{NP}\, \mu}\simeq -0.73$ or  $ C_9^{\mathrm{NP}\, \mu}= -  C_{10}^{\mathrm{NP}\, \mu}\simeq -0.39$,  can lead to a significant improvement of the quality of the fit, compared to the SM-only hypothesis. Mainstream models are loop-induced new physics, as well as
tree-level new physics contributions, mediated by  $Z^\prime$ gauge bosons or leptoquarks. At a variance with the $b \to c$ anomalies, the new physics scale is found of order 40 TeV, so one cannot check direct production of new particles at the (HL-)LHC.
There is  a limited reach for searches for deviations from the SM in high $p_T$ di-muon tails.
The couplings are stringently constrained by $B^0_s$ meson mixing data. One has also to remark that vector leptoquarks can provide a unique solution for both  $ b\to c$ and $b \to$ s anomalies.
Last but not least, 
the  discrepancies with SM predictions have also generated  alternative data analyses, as for example the  Dispersive Matrix method, claiming  the values $R(D)=0.296 \pm 0.008$ and $R(D^\ast)=0.261 \pm 0.020$, which differ by  about 1.4$\sigma$ from the latest experimental determinations \cite{PhysRevD.105.034503}.

\section{$|V_{xb}|$  determination}

 Improving our knowledge of $|V_{cb}|$ and $|V_{ub}|$ and reducing their uncertainty
are crucial  to the so-called  $|V_{cb}|$ and $|V_{ub}|$ puzzles, the longstanding tension between the values obtained from exclusive  and inclusive  determinations..
The latest global analysis of the inclusive $B\to X_c\ell\nu$  by HFLAV \cite{HFLAV:2019otj}
gives $
|V_{cb}^\mathrm{incl}|=(42.19\pm 0.78)\times 10^{-3}
$. It has been done
in  the framework of kinetic scheme, where $|V_{cb}|$ is extracted together with the $b$ and $c$ quark masses and 4 non-perturbative  parameters (namely $\mu^2_{\pi}$, $\mu^2_{G}$,  $\rho^3_{D}$ and $\rho^3_{LS}$). Details on the extraction can be found for instance in Ref. \cite{Ricciardi:2019zph}.
More recent results  are generally consistent with the HFLAV analysis,  as \cite{Bordone:2021oof}
$
|V_{cb}^\mathrm{incl}|=(42.16\pm 0.50)\times 10^{-3}
$,  or the first determination  using moments
of the dilepton invariant mass \cite{Bernlochner:2022ucr}
$
|V_{cb}^\mathrm{incl}|=(41.69\pm 0.63)\times 10^{-3}
$. 
One can compare these results with the latest FLAG determination for exclusive decays \cite{FlavourLatticeAveragingGroupFLAG:2021npn}
$
|V_{cb}^\mathrm{excl}|=(39.36\pm 0.68)\times 10^{-3}
$. In the case of Ref. \cite{Bordone:2021oof}, for instance, one finds a 2.6$\sigma$ difference. 
At a variance with previous determinations, in the dispersion matrix
approach \cite{Martinelli:2022skp} the $|V_{cb}|$ inclusive/exclusive values are compatible within the 1$\sigma$ level. Novel approaches have been recently put forward, that allow one to address
inclusive decays in LQCD, see for instance Ref. \cite{Gambino:2022dvu} and references therein.

In order to extract  $|V_{ub}|$  from semileptonic $B \to X_u \ell \nu$ decays one has to  reduce the $b \to c$ semileptonic background through experimental  cuts. Such cuts enhance the relevance of the so-called threshold region in the phase space, jeopardizing the use of the heavy quark expansion. In order to face this problem, that is absent in the inclusive determination of $|V_{cb}|$,   different theoretical schemes have been devised, which are  tailored
to analyze data in the threshold region,  but  differ
in their treatment of perturbative corrections and the
parametrization of non-perturbative effects.
The value of  $|V_{ub}|$ has been extracted by BaBar \cite{Lees:2011fv, Beleno:2013jla}, Belle \cite{Urquijo:2009tp,Belle:2021ymg, Cao:2021uwy}  and  HFLAV  \cite{Amhis:2016xyh} collaborations. It is based on the measured partial branching fractions value with  the state-of-the-art theory predictions on decay rate in 
four different theoretical different approaches: ADFR  \cite{Aglietti:2004fz, Aglietti:2006yb,  Aglietti:2007ik}, BLNP
 \cite{Lange:2005yw, Bosch:2004th, Bosch:2004cb}, DGE \cite{Andersen:2005mj} and  GGOU  \cite{Gambino:2007rp}.
 The arithmetic average of
the most precise determinations for the phase-space region $E_\ell ^B >1$
gives (Belle 2022 \cite{Cao:2021qpa})   
$
|V_{ub}^\mathrm{incl}|=(4.10 \pm 0.09\pm 0.22 \pm 0.15) \times 10^{-3}
$,
where the  uncertainties are
statistical, systematic and  from the  theory calculation, respectively.
This can be compared with the world average of exclusive
results 
$
|V_{ub}^\mathrm{excl}|=(3.67 \pm 0.09\pm 0.12) \times 10^{-3}
$.
The 2022 Belle inclusive value is smaller than the previous inclusive ones,  reducing the discrepancy with the exclusive measurement of about 2-3 to  1.3 standard deviations. It is also compatible with
the value expected from CKM unitarity from a global fit \cite{Charles:2004jd}
$
|V_{ub}|=(3.62^{+0.11}_{-0.08})  \times 10^{-3}
 $ within 1.6 standard deviations.






\bibliography{VxbRef}
%

\end{document}